\documentclass[12pt]{article}
\usepackage{epsfig}
\usepackage{multirow}

\newcommand{\mysection}{\setcounter{equation}{0}\section}

\def\beq{\begin{equation}}
\def\eeq{\end{equation}}
\def\beqa{\begin{eqnarray}}
\def\eeqa{\end{eqnarray}}
 
\newlength{\dinwidth} \newlength{\dinmargin}
\begin{document}

\begin{center}
{\Large \bf NNNLO soft-gluon corrections for the top-quark $p_T$ and rapidity distributions}
\end{center}
\vspace{2mm}
\begin{center}
{\large Nikolaos Kidonakis}\\
\vspace{2mm}
{\it Department of Physics, Kennesaw State University, \\
Kennesaw, GA 30144, USA}
\end{center}
 
\begin{abstract}
I present a calculation of next-to-next-to-next-to-leading-order (NNNLO) 
soft-gluon corrections for differential distributions in top-antitop pair 
production in hadronic collisions. 
Approximate NNNLO (aNNNLO) results are obtained from soft-gluon resummation. 
Theoretical predictions are shown for the top-quark aNNNLO transverse 
momentum ($p_T$) and 
rapidity distributions at LHC and Tevatron energies. The aNNNLO corrections 
enhance previous results for the distributions 
but have smaller theoretical uncertainties.
\end{abstract}
 
\mysection{Introduction}

Top quark physics occupies a central role in particle theory and experiment. The top quark has a unique position as the most massive elementary particle to have been discovered, and as the only quark that decays before it can hadronize. Thus understanding the properties and production rates of the top quark in the Standard Model is crucial for QCD, electroweak and Higgs physics, and in searches for new physics.

The production of top quarks can happen either via top-antitop pair production or via single-top production. While the total cross section is the simplest quantity to be calculated and measured, differential distributions provide more detailed information on the production processes. The top-quark transverse momentum ($p_T$) and rapidity distributions are particularly useful for discriminating signals of new physics. They have been measured at the Tevatron \cite{CDF,D0} and the LHC \cite{ATLAS,CMS} and are in excellent agreement with approximate NNLO (aNNLO) calculations for the $p_T$ \cite{NKpt} and rapidity \cite{NKy} distributions (see also \cite{NKHQ13} for updates). These aNNLO calculations are based on the resummation of soft-gluon contributions in the double-differential cross section at next-to-next-to-leading-logarithm (NNLL) accuracy in the moment-space approach in perturbative QCD \cite{NKpt,NKy,NKHQ13}. For a review of resummation in various approaches for top quark production see the review paper in \cite{NKBP}. The theoretical and experimental status of top quark physics has also been more recently reviewed in \cite{FJS,TopWG,SJ,GV}.  

The increasing precision of the measurements at the LHC and the upcoming run at 13 TeV necessitate ever more precise theoretical calculations. For the total cross section the current state-of-the-art is approximate NNNLO (aNNNLO) \cite{NKaNNNLO}. The purpose of the present paper is to bring the calculation of the top-quark differential distributions to the same accuracy as the total cross section in \cite{NKaNNNLO}. The resummation formalism for $t{\bar t}$ production has already been presented and discussed extensively in Refs. [5-8,13,14] and we refer the reader to those papers for more details and further references. The accuracy, stability, and reliability of the resummation formalism that we use has been amply demonstrated and discussed in \cite{NKpt,NKHQ13}; the soft-gluon corrections approximate exact results for total cross sections and differential distributions at the per mille accuracy level.  

In Section 2 we present some details of the formalism and kinematics. 
In Section 3 we present the top-quark $p_T$ distributions at LHC and Tevatron energies. Section 4 contains the corresponding top-quark rapidity distributions.
We conclude in Section 5.

\mysection{Soft-gluon corrections and kinematics for differential distributions}

At leading order in the strong coupling, $\alpha_s$,  
for $t{\bar t}$ production, i.e. ${\cal O}(\alpha_s^2)$, 
the partonic channels are quark-antiquark annihilation, 
\beq
q(p_1)+{\bar q}(p_2) \rightarrow t(p_3) +{\bar t}(p_4) \, , 
\eeq
and gluon-gluon fusion,
\beq
g(p_1)+g(p_2) \rightarrow t(p_3) +{\bar t}(p_4) \, .
\eeq

We define the partonic variables $s=(p_1+p_2)^2$,  $t_1=(p_1-p_3)^2-m_t^2$, 
$u_1=(p_2-p_3)^2-m_t^2$, and $\beta=\sqrt{1-4m_t^2/s}$, 
where $m_t$ is the top-quark mass. 
We also define the threshold quantity $s_4=s+t_1+u_1$. At partonic threshold 
there is no energy available for additional radiation and $s_4$ vanishes in 
that limit but the top quark can have arbitrarily large  
transverse momentum or rapidity. Near partonic threshold the gluons radiated 
are soft, i.e. have low energy.

Soft-gluon corrections appear in the partonic cross section  
as plus distributions of logarithms of $s_4$, which are  
defined through their integral with parton distribution functions, $\phi$, as 
\beqa
\int_0^{s_{4 \, \rm{max}}} ds_4 \, \phi(s_4) \left[\frac{\ln^k(s_4/m_t^2)}
{s_4}\right]_{+} &\equiv&
\int_0^{s_{4\, {\rm max}}} ds_4 \frac{\ln^k(s_4/m_t^2)}{s_4} [\phi(s_4) - \phi(0)]
\nonumber \\ &&
{}+\frac{1}{k+1} \ln^{k+1}\left(\frac{s_{4\, {\rm max}}}{m_t^2}\right) \phi(0) 
\label{plusd}
\eeqa
where the integer $k$ ranges from 0 to $2n-1$ for the $n$th order corrections, 
i.e. the ${\cal O}(\alpha_s^{2+n})$ terms.
The derivation of these corrections follows from renormalization group evolution of the soft and jet functions in the factorized cross section \cite{NKpt,NKHQ13,NKBP}.

At NNNLO the soft-gluon corrections to the double-differential partonic 
cross section, $d^2{\hat \sigma}/(dt_1 \,du_1)$, take the form 
\beq 
\frac{d^2{\hat \sigma}^{(3)}}{dt_1 \, du_1}=\alpha_s^5 
\sum_{k=0}^5 C_k^{(3)}(s_4) \left[\frac{\ln^k(s_4/m_t^2)}{s_4}\right]_+ \, .
\eeq
The coefficients $C_k^{(3)}$ are in general functions of $s_4$, 
$s$, $t$, $u$, $m_t$, the renormalization scale $\mu_R$, 
and the factorization scale $\mu_F$; 
these coefficients have been determined from NNLL resummation 
\cite{NKaNNNLO,NKNNNLO} and the expressions beyond the leading logarithms 
(i.e. below $k=5$) are long.
For simplicity we only show the dependence on $s_4$ in the argument of the 
$C_k^{(3)}$ as it will be needed explicitly below.

The hadronic variables for the process with incoming hadrons $h_1$ and $h_2$  
(protons at the LHC; protons and antiprotons at the Tevatron),
\beq
h_1(p_{h_1})+h_2(p_{h_2}) \rightarrow t(p_3) +{\bar t}(p_4) \, ,
\eeq
are 
$S=(p_{h1}+p_{h2})^2$,  $T_1=(p_{h1}-p_3)^2-m_t^2$, and $U_1=(p_{h2}-p_3)^2-m_t^2$.
Note that $p_1=x_1 p_{h1}$ and $p_2=x_2 p_{h2}$, where $x_i$ is the parton 
momentum fraction in hadron $h_i$, and thus $t_1=x_1 T_1$, 
$u_1=x_2 U_1$, and $s=x_1 x_2 S$.
The top-quark transverse momentum is 
$p_T=|{\bf p}_3| \sin\theta$, with $\theta$ the angle between ${\bf p}_{h_1}$
and ${\bf p}_3$, while the top-quark rapidity is  
\beq
Y=\frac{1}{2}\ln\left(\frac{E_3+p_{3z}}{E_3-p_{3z}}\right)
\eeq
where $p_{3z}=|{\bf p}_3| \cos\theta$.
Note the relations
\beq
p_T^2=\frac{T_1 \, U_1}{S}-m_t^2=\frac{t_1 \, u_1}{s}-m_t^2 \, ,
\eeq
\beq
Y=\frac{1}{2}\ln\left(\frac{U_1}{T_1}\right) \, ,
\eeq
$U_1=-\sqrt{S} m_T e^Y$, and $T_1=-\sqrt{S} m_T e^{-Y}$, 
with $m_T=\sqrt{m_t^2+p_T^2}$. 

The relation between $x_2$ and the other variables is 
\beq
x_2=\frac{s_4-x_1 T_1}{x_1S+U_1} \, .
\eeq
The elastic expression for $x_2$ , i.e. when $s_4=0$, is
\beq
x_2^{\rm el}=\frac{-x_1 T_1}{x_1S+U_1} \, .
\eeq

Then we can write the hadronic aNNNLO corrections for the double-differential cross section in $p_T$ and rapidity as
\beq
\frac{d^2\sigma^{(3)}}{dp_T^2 \, dY}= 
\alpha_s^5 
\int_{x_1^-}^1 dx_1  
\int_0^{s_{4 \, {\rm max}}} ds_4 \frac{x_1 x_2 S}{x_1 S+ U_1} \, 
\phi(x_1) \phi(x_2) 
\sum_{k=0}^{5}
C_k^{(3)}(s_4) \left[\frac{\ln^k(s_4/m_t^2)}{s_4}\right]_+  \, , 
\eeq
where
$x_1^-=-U_1/(S+T_1)$ and $s_{4 \, {\rm max}}=x_1(S+T_1)+U_1$.

We can rewrite the above expression, using Eq. (\ref{plusd}),  as
\beqa
&& \hspace{-12mm}\frac{d^2\sigma^{(3)}}{dp_T^2 \, dY}= 
\alpha_s^5 \int_{x_1^-}^1 dx_1 \frac{x_1 S}{x_1 S+ U_1} \phi(x_1)
\sum_{k=0}^5 \left\{\int_0^{s_{4 \, {\rm max}}} \frac{ds_4}{s_4} 
\ln^k\left(\frac{s_4}{m_t^2}\right) 
\left[C_k^{(3)}(s_4) x_2 \phi(x_2)
-C_k^{(3)}(0) x_2^{\rm el} \phi(x_2^{\rm el})\right] \right.
\nonumber \\ && \hspace{58mm} \left. 
{}+\frac{1}{k+1}
\ln^{k+1}\left(\frac{s_{4 \, {\rm max}}}{m_t^2}\right) C_k^{(3)}(0)  
x_2^{\rm el} \phi(x_2^{\rm el}) \right\} \, .
\eeqa

The aNNNLO correction to the top quark transverse momentum distribution is 
then given by
\beqa
\frac{d\sigma^{(3)}}{dp_T}&=& 
2 p_T \int_{Y^-}^{Y^+} \frac{d^2\sigma^{(3)}}{dp_T^2 \, dY} dY 
\eeqa
where
\beq
Y^{\pm}=\pm \frac{1}{2} \ln \frac{1+\sqrt{1-\frac{4m_T^2}{S}}}
{1-\sqrt{1-\frac{4m_T^2}{S}}} \, .
\eeq

The aNNNLO correction to the top quark rapidity distribution is given by 
\beqa
\frac{d\sigma^{(3)}}{dY}&=& 
\int_0^{p_T^{+\, 2}} \frac{d^2\sigma^{(3)}}{dp_T^2 \, dY} dp_T^2 
\eeqa
where
\beq
p_{T}^{+\, 2}=\frac{S}{4\cosh^2Y}-m_t^2 \, .
\eeq

Note that the total cross section can be simply found by integrating either 
over the $p_T$ distribution
\beq
\sigma=\int_0^{p_{T\, \rm max}} \frac{d\sigma}{dp_T} \, dp_T
\eeq
with $p_{T\, \rm max}=\sqrt{\frac{S}{4}-m_t^2}$; 
or over the rapidity distribution
\beq
\sigma=\int_{Y_{\rm min}}^{Y^{\rm \max}} \frac{d\sigma}{dY} \, dY
\eeq
with
\beq
Y^{\rm max}_{\rm min}=\pm \frac{1}{2} \ln\left(\frac{1+\beta_h}{1-\beta_h}\right)
\eeq
and $\beta_h=\sqrt{1-\frac{4m_t^2}{S}}$ the hadronic analog of $\beta$.

\mysection{Top-quark transverse momentum distributions}

We use the expressions in Section 2 to calculate the top-quark differential 
distributions at aNNNLO.
We begin with the top-quark transverse momentum distributions.
We use the MSTW2008 NNLO parton distribution functions (pdf) \cite{MSTW} 
for all our numerical results.

\begin{figure}
\begin{center}
\includegraphics[width=10cm]{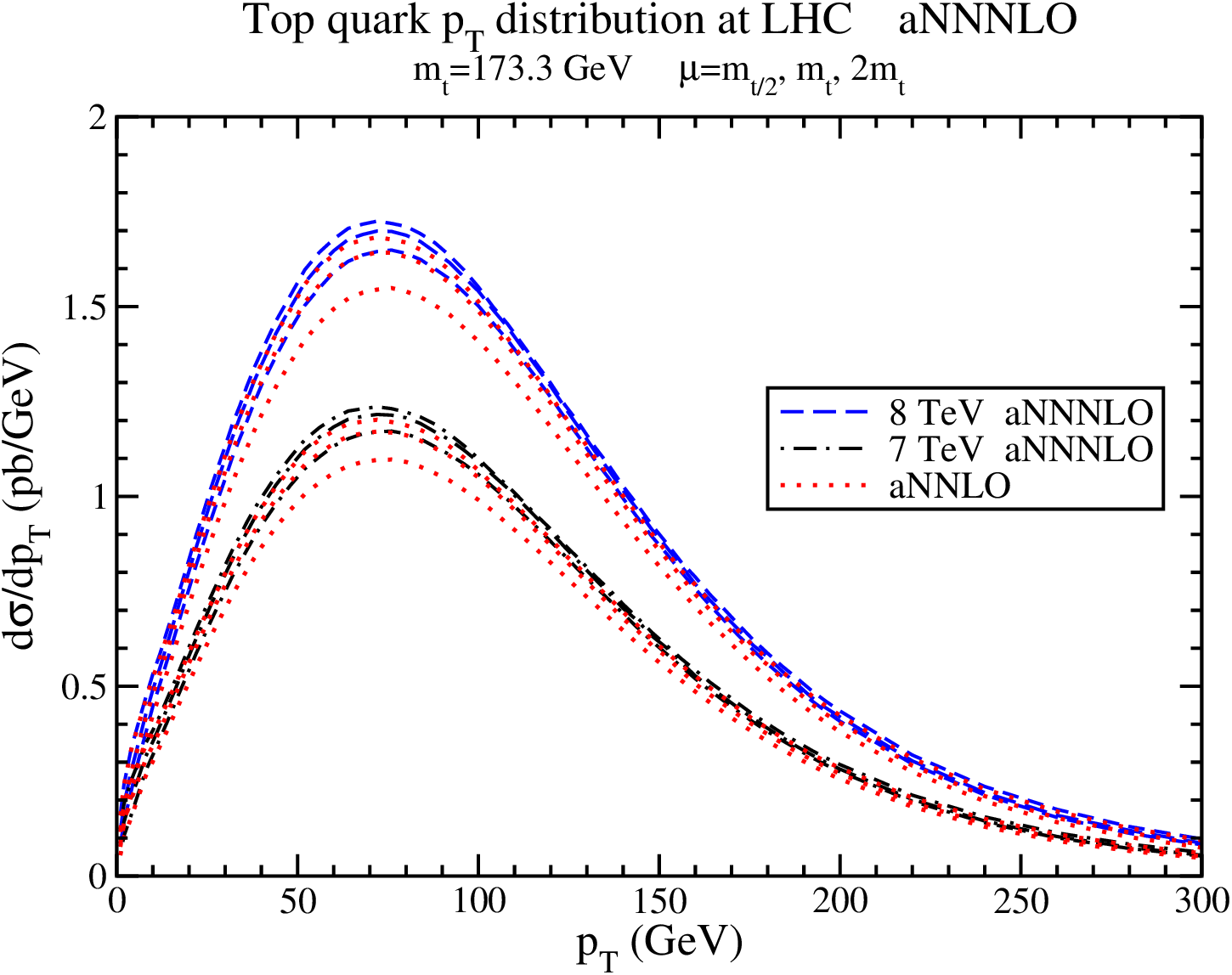}
\caption{The aNNNLO top-quark $p_T$ distributions in $t{\bar t}$ production 
at the LHC with $\sqrt{S}=7$ and 8 TeV.}
\label{pttop7-8lhc}
\end{center}
\end{figure}

In Fig. \ref{pttop7-8lhc} we show the aNNNLO top-quark $p_T$ distributions 
at 7 and 8 TeV LHC energy for $p_T$ values up to 300 GeV. For comparison 
we also show the aNNLO results at each energy.
The central lines are with scale $\mu=m_t$, and the other lines display 
the upper and lower values from scale 
variation over the interval $m_t/2 \le \mu \le 2m_t$ for each energy.
The uncertainty from scale variation at aNNNLO is smaller than at 
aNNLO, a fact which is consistent with the results for the total 
cross section in \cite{NKaNNNLO}. 

\begin{figure}
\begin{center}
\includegraphics[width=10cm]{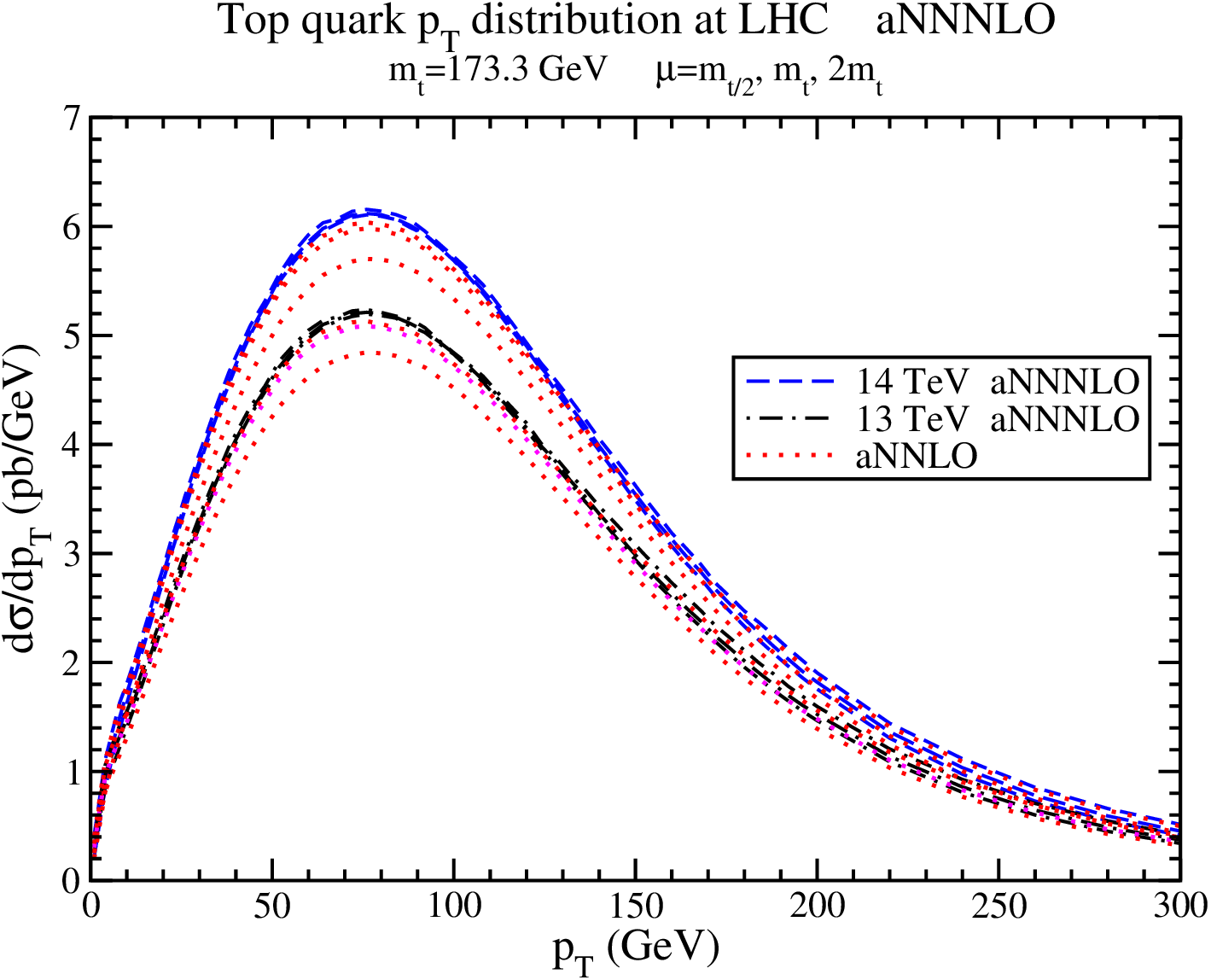}
\caption{The aNNNLO top-quark $p_T$ distributions in 
$t{\bar t}$ production at the LHC with $\sqrt{S}=13$ and 14 TeV.}
\label{pttop13-14lhc}
\end{center}
\end{figure}

In Fig. \ref{pttop13-14lhc} we show the aNNNLO top-quark $p_T$ distributions 
at 13 and 14 TeV LHC energy, also for $p_T$ values up to 300 GeV. 
Again, for comparison, we also display the aNNLO curves for each energy. The 
scale variation is again smaller at aNNNLO than at lower orders.  

\begin{figure}
\begin{center}
\includegraphics[width=10cm]{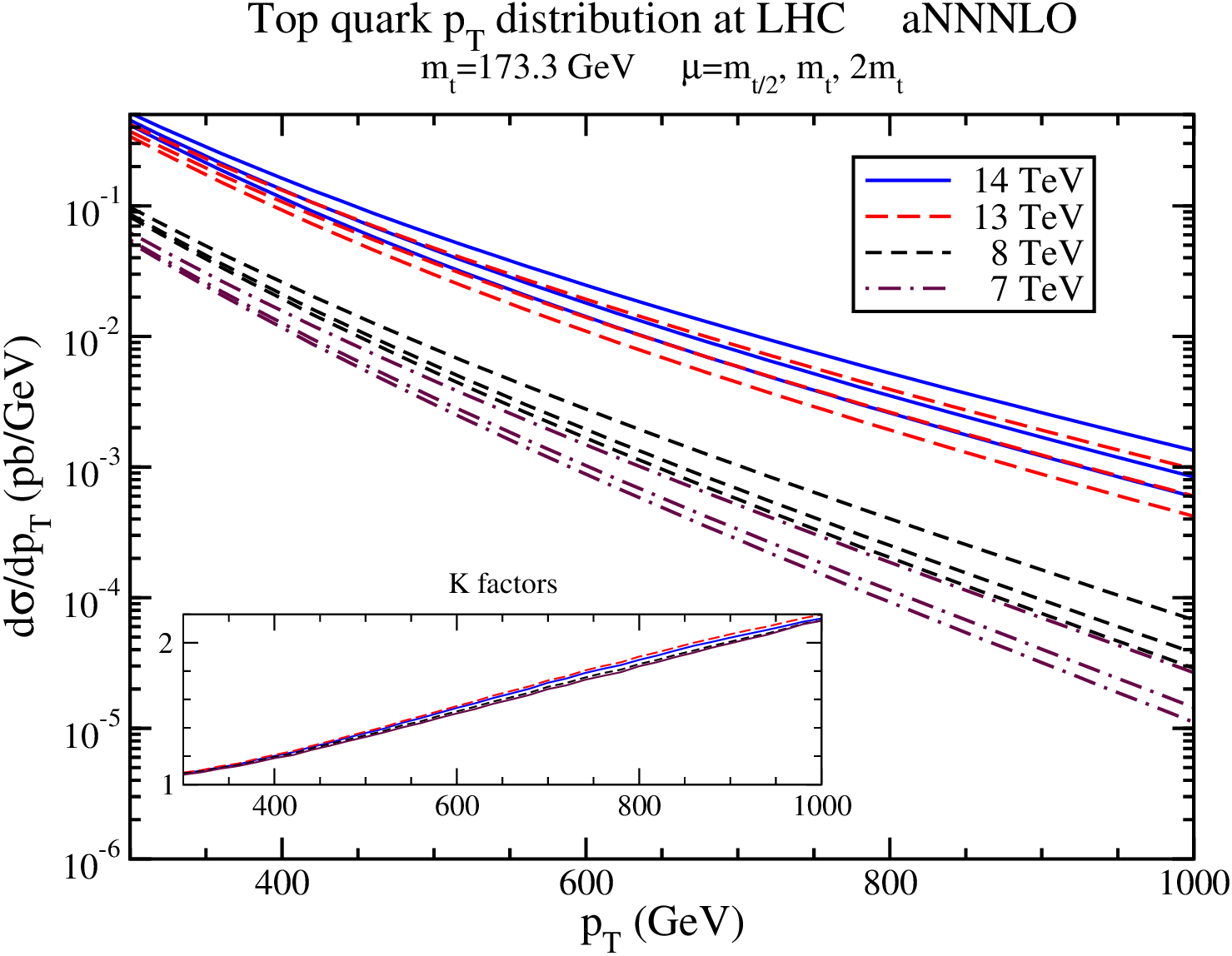}
\caption{The aNNNLO top-quark $p_T$ distributions at high $p_T$ in 
$t{\bar t}$ production at the LHC with  $\sqrt{S}=7$, 8, 13, and 14 TeV.  
The inset plot shows the $K$ factors.}
\label{pttoplhclog}
\end{center}
\end{figure}

Figure \ref{pttoplhclog} displays the aNNNLO top-quark 
$p_T$ distributions at high $p_T$ at LHC energies in a logarithmic plot over 
a larger range of transverse momenta, 300 GeV $\le p_T \le 1000$ GeV. 
The distributions span 4 to 5 orders of magnitude over the $p_T$ range shown 
in the figures. 
The inset plot displays  the $K$-factors, i.e. the ratios of the aNNNLO 
$p_T$ distribution to the NLO \cite{NLO1,NLO2} result, 
for each energy, all with the same choice 
of pdf. We observe that the $K$-factors increase with increasing $p_T$ 
as would be expected since we get closer to partonic threshold.

\begin{figure}
\begin{center}
\includegraphics[width=10cm]{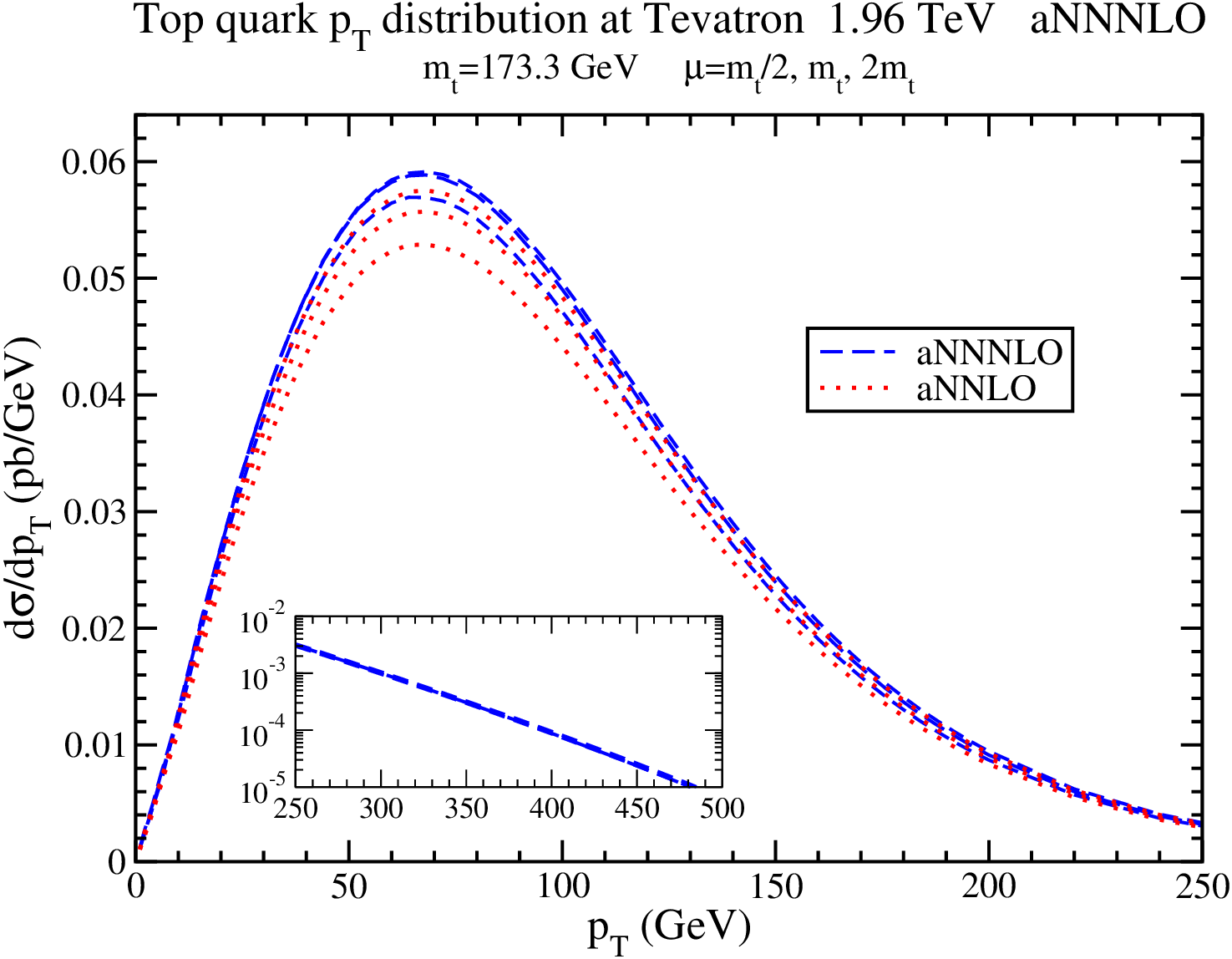}
\caption{The aNNNLO top-quark $p_T$ distributions in 
$t{\bar t}$ production at the Tevatron  with $\sqrt{S}=1.96$ TeV.}
\label{pttoptev}
\end{center}
\end{figure}

Figure \ref{pttoptev} displays the aNNNLO top-quark 
$p_T$ distributions (and also the aNNLO results) at the Tevatron with 
1.96 TeV energy. Again, the uncertainty from scale variation at aNNNLO is 
smaller than in previous calculations. The inset plot displays 
the high-$p_T$ region in a logarithmic scale up to a $p_T$ of 500 GeV.

As a numerical consistency check, we note that the aNNNLO cross sections found 
by integrating over the $p_T$ distributions in Figs. 1-4 agree with the 
results in \cite{NKaNNNLO}.

\mysection{Top-quark rapidity distributions}

We continue with the aNNNLO top-quark rapidity distributions 
in $t{\bar t}$ production at the LHC and the Tevatron.

\begin{figure}
\begin{center}
\includegraphics[width=10cm]{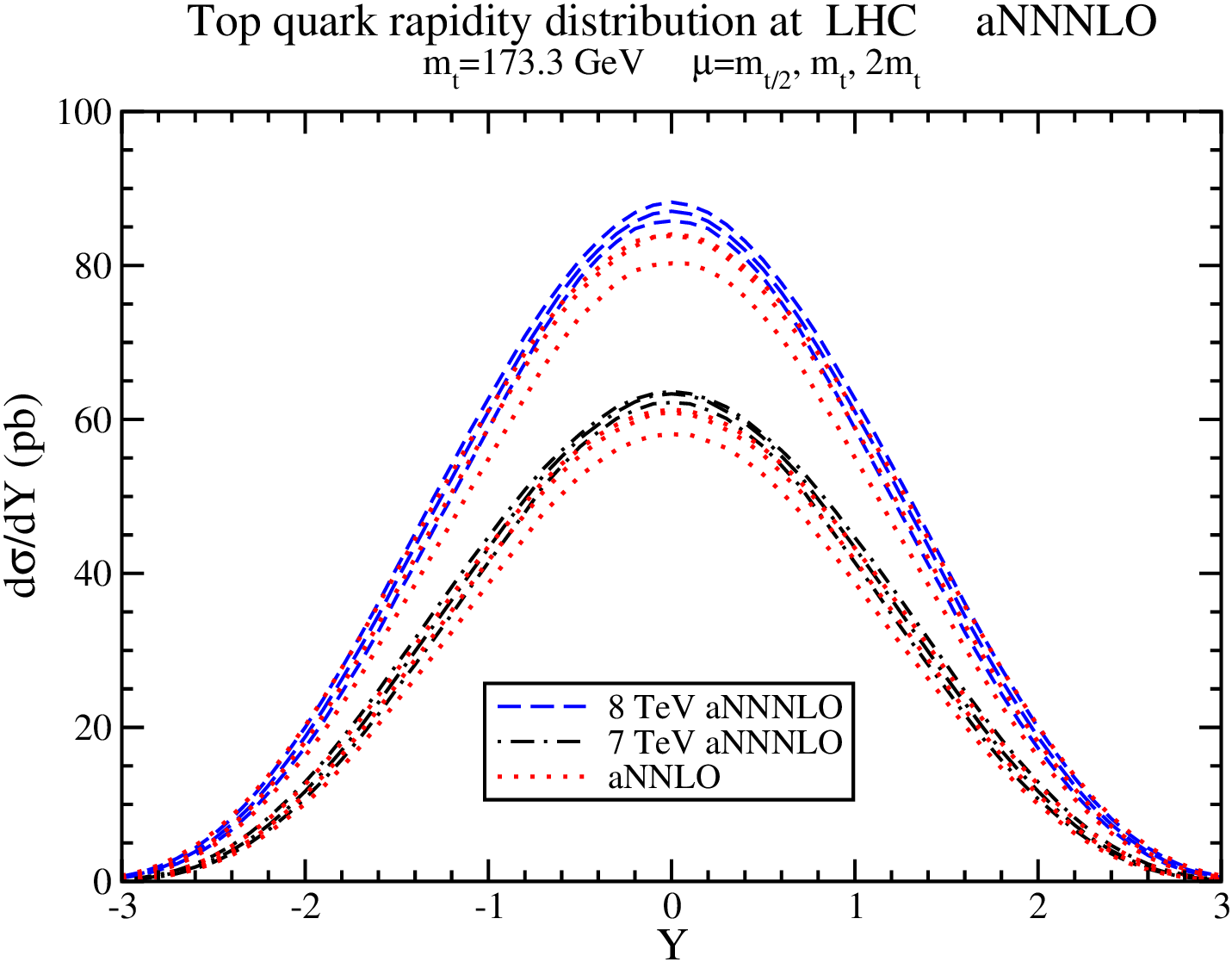}
\caption{The aNNNLO top-quark rapidity distributions 
in $t{\bar t}$ production at the LHC with $\sqrt{S}=7$ and 8 TeV.}
\label{ytop7-8lhc}
\end{center}
\end{figure}

In Fig. \ref{ytop7-8lhc} we show the aNNNLO top-quark rapidity distributions 
at 7 and 8 TeV LHC energy. We also show the aNNLO results at each energy.
The central lines are with 
$\mu=m_t$, and the other lines display the upper and lower values from scale 
variation over the interval $m_t/2 \le \mu \le 2m_t$ for each energy.
The uncertainty from scale variation at aNNNLO is smaller than at 
aNNLO, a fact which is consistent with the results for the total 
cross section in \cite{NKaNNNLO} as well as the observations in the previous 
section. 

\begin{figure}
\begin{center}
\includegraphics[width=10cm]{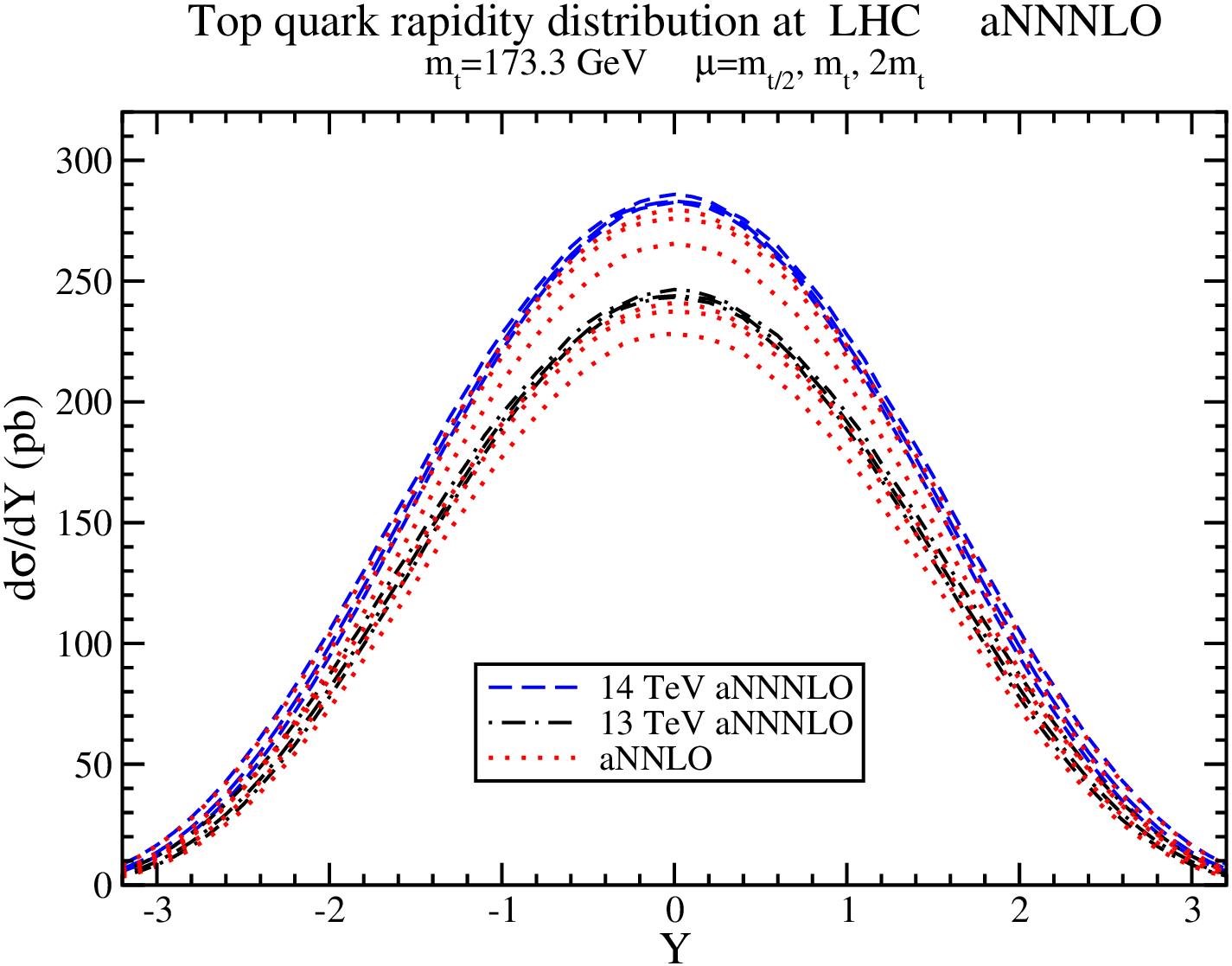}
\caption{The aNNNLO top-quark rapidity distributions 
in $t{\bar t}$ production at the LHC with $\sqrt{S}=13$ and 14 TeV.}
\label{ytop13-14lhc}
\end{center}
\end{figure}

In Fig. \ref{ytop13-14lhc} we show the aNNNLO (and also the aNNLO) 
top-quark rapidity distributions 
at 13 and 14 TeV LHC energy. Again, the scale variation at aNNNLO  
is smaller than that at aNNLO.

\begin{figure}
\begin{center}
\includegraphics[width=10cm]{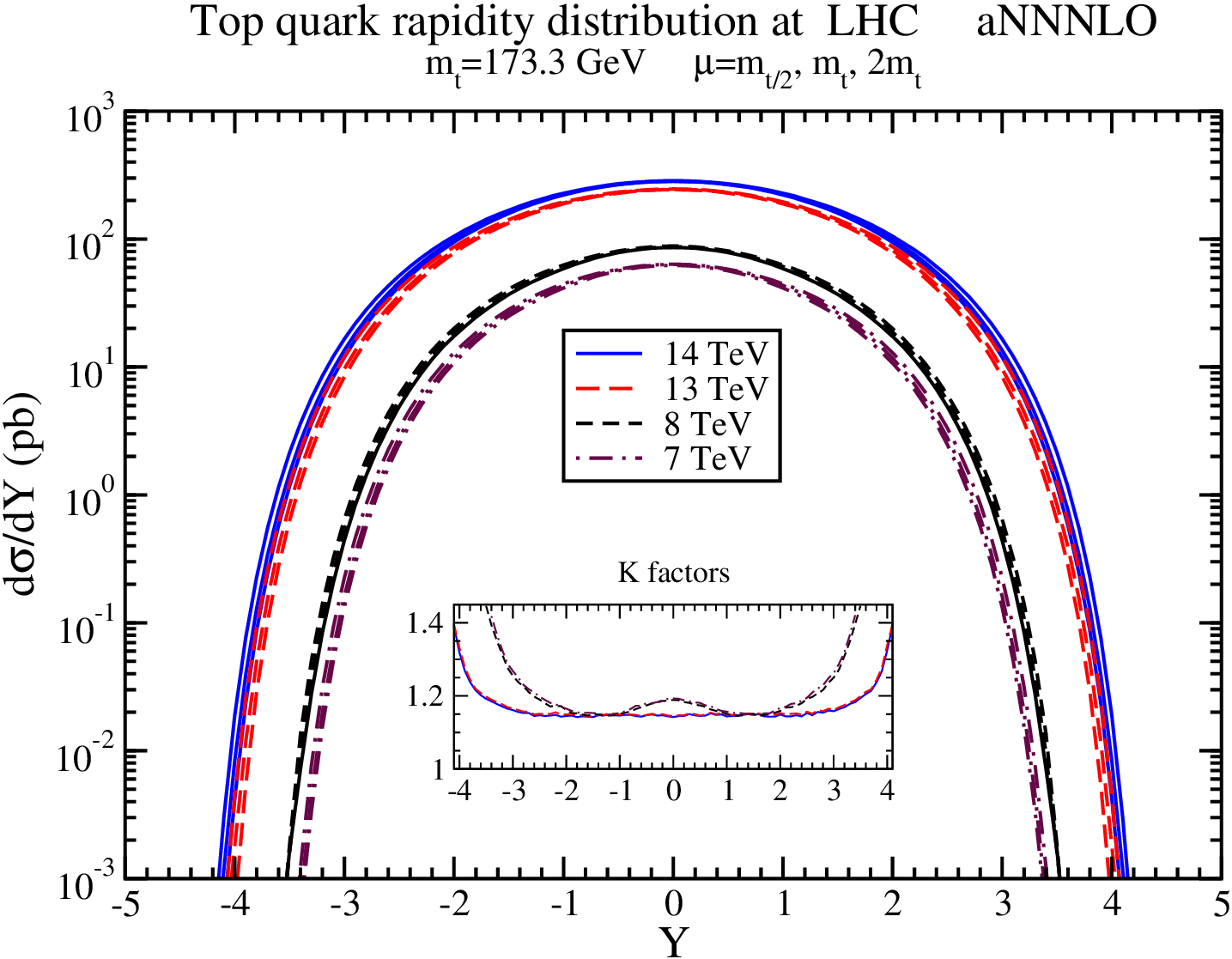}
\caption{The aNNNLO top-quark rapidity distributions 
in $t{\bar t}$ production at the LHC with $\sqrt{S}=7$, 8, 13, and 14 TeV. 
The inset plot shows the $K$ factors.}
\label{ytoplhclog}
\end{center}
\end{figure}

Figure \ref{ytoplhclog} displays the aNNNLO top-quark 
rapidity distributions at LHC 
energies in a logarithmic plot over a larger range of rapidities. 
The inset plot displays  the $K$-factors, i.e. the ratios of the aNNNLO 
rapidity distribution to the NLO \cite{NLO1,NLO2} result, for each energy,  
all with the same choice of pdf. The $K$-factors are larger at the edges 
of the rapidity distribution, i.e. at large absolute value of the rapidity, 
as expected.

\begin{figure}
\begin{center}
\includegraphics[width=10cm]{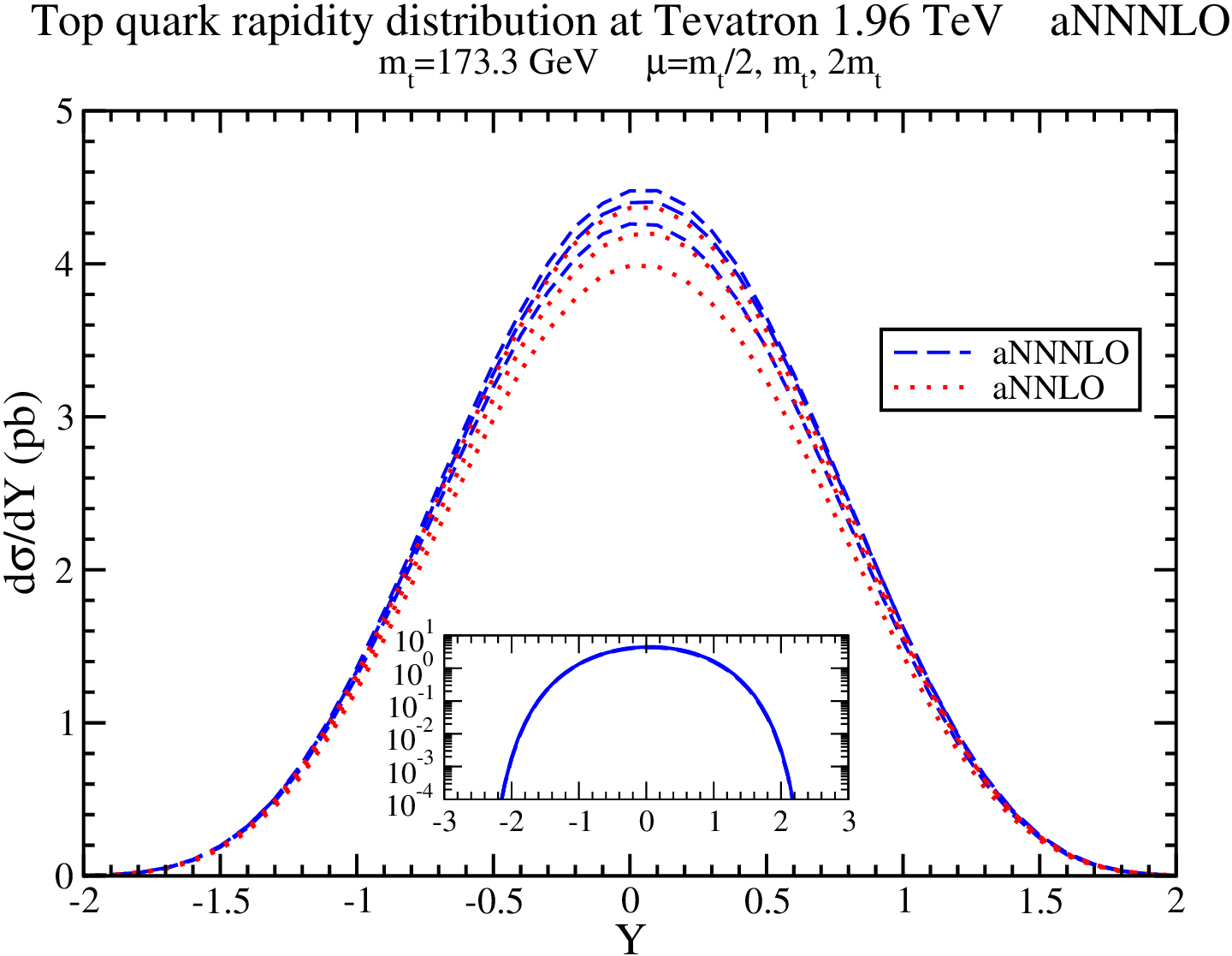}
\caption{The aNNNLO top-quark rapidity distributions 
in $t{\bar t}$ production at the Tevatron with $\sqrt{S}=1.96$ TeV.}
\label{ytoptev}
\end{center}
\end{figure}

Figure \ref{ytoptev} displays the aNNNLO top-quark rapidity distributions at 
the Tevatron at 1.96 TeV energy and, for comparison, also the aNNLO results.  
 Again, the scale variation at aNNNLO is 
smaller than in previous calculations. 
The inset plot displays a wider rapidity region in a logarithmic scale.

As before, as a numerical consistency check, we note that the aNNNLO 
cross sections found by integrating over the rapidity distributions in 
Figs. 5-8 agree with the results in \cite{NKaNNNLO}.

\mysection{Conclusions}

The aNNNLO top-quark transverse momentum and rapidity distributions have been calculated in $t{\bar t}$ production at the LHC at 7, 8, 13, and 14 TeV energies, and at the Tevatron at 1.96 TeV energy. The distributions enhance previous results but they reduce the theoretical uncertainty. The enhancement is particularly large at very high $p_T$ for the transverse momentum distributions and at large absolute value of the rapidity for the rapidity distributions.

\mysection*{Acknowledgements}
This material is based upon work supported by the National Science Foundation 
under Grant No. PHY 1212472.

\end{document}